\documentclass[useAMS]{mn2e}
\usepackage{epsfig}
\title[Dark halo spin parameter]
{The Evolution of the Dark Halo Spin Parameters $\lambda$ and
$\lambda'$ in a $\Lambda$CDM Universe: The Role of Minor and Major Mergers}
\author[Helmut Hetznecker \& Andreas Burkert]
{Helmut Hetznecker$^{1}$\thanks{E-mail:hetzneck@mpia.de},
Andreas Burkert$^2$\thanks{E-mail:burkert@usm.uni-muenchen.de}\\
$^{1}$Max-Planck-Institut f\"ur Astronomie,
K\"onigstuhl 17, D-69117 Heidelberg, Germany, \\
$^2$University Observatory Munich,
Scheinerstr.~1, D-81679, M\"unchen, Germany}
\date{Submitted to MNRAS}

\pagerange{\pageref{firstpage}--\pageref{lastpage}} \pubyear{2005}

\def\LaTeX{L\kern-.36em\raise.3ex\hbox{a}\kern-.15em
    T\kern-.1667em\lower.7ex\hbox{E}\kern-.125emX}

\newcommand{\msol}{M_\odot}

\newcommand{\beq}{\begin{equation}}
\newcommand{\eeq}{\end{equation}}

\newcommand{\rp}{R_{\rm peri}}
\newcommand{\rv}{R_{\rm vir}}
\newcommand{\li}{\lambda_{\rm{init}}}
\newcommand{\lf}{\lambda_{\rm{fin}}}
\newcommand{\lp}{\lambda^{(\prime)}}
\newcommand{\tl}{T_{\lambda}}
\newcommand{\tlp}{T_{\lambda}^{(\prime)}}
\newcommand{\oli}{\overline}
\newcommand{\lzm}{\overline\lambda(z)}
\newcommand{\ovl}{\overline\lambda}
\newcommand{\ek}{E_{\rm kin}}
\newcommand{\ep}{E_{\rm pot}}
\newcommand{\oeta}{\overline\eta}
\newcommand{\lovl}{\lambda'/\lambda}

\begin{document}

\label{firstpage}

\maketitle

\begin{abstract}

The evolution of the spin parameter of dark halos and the dependence on the 
halo merging history in a set of dissipationless cosmological $\Lambda$CDM 
simulations is investigated. Special focus is placed on the differences of 
the two commonly used versions
of the spin parameter, namely $\lambda=JE^{1/2}G^{-1}M^{-5/2}$ (Peebles 1980) 
and $\lambda'=J/[\sqrt{2}M_{\rm vir}R_{\rm vir}V_{\rm vir}]$ (Bullock et al. 2001). 
Though the distribution of the spin transfer rate 
$\tlp:=\lf^{(\prime)}/\li^{(\prime)}$ (which is the ratio of spin parameters
after and prior to merger) is similar to a high degree for both
$\lambda$ and $\lambda'$, we find considerable differences in the time 
evolution: while $\lambda'$ is roughly independent of redshift, $\lambda$ 
turns out to increase significantly with decreasing redshift. This distinct 
behaviour arises from small differences in the spin transfer distribution 
of accreted material. 
The evolution of the spin parameter is strongly coupled with the virial ratio 
$\eta:=2\ek/|\ep|$ of dark halos. Major mergers disturb halos and increase both 
their virial ratio $\eta$ and spin parameter $\lp$ for $1-2$ Gyrs. At high
redshifts $z=2-3$ many halos are disturbed with an average virial ratio of
$\oli\eta\approx 1.3$ which approaches unity until $z=0$. We find that the
redshift evolution of the spin parameters is dominated by the huge number of 
minor mergers rather than the rare major merger events.

\end{abstract}


\section{Introduction}

It is among the puzzling tasks within the theory of galaxy formation to 
understand the origin and history of angular momentum both of dark matter 
halos and embedded galactic disks. One reason for the high relevance of galactic spin 
is that it determines the most basic properties of a galactic disc, its 
surface density distribution and its spatial extension through the 
equilibrium  of gravitational and centrifugal forces, as discussed already 
by Hoyle as early as 1949 (Hoyle 1949). 
A kind of basic picture of galactic evolution 
has emerged in the undertow of the seminal works of White \& Rees (1978) and 
Fall \& Efstathiou (1980), which connects the angular momentum of a dark halo to  
its mass growth history (see also Blumenthal et al. 1986, Mo, Mao \& White 1998,
van den Bosch 2000, Cole et al. 2000, Somerville, Primack \& Faber 2001, Kauffmann, White
\& Guiderdoni 1993).

Until the beginning of this century it was considered as established that 
bound cosmological structures gain most of their angular momentum 
by the tidal effect of the large scale mass distribution: After a merger the
remnant is left with the angular momentum imprint of the tidal field. 
 In recent years it was realized that yet another effect is contributing to 
a halo's angular momentum history to at least the same degree as large scale 
tidal forces, namely its discrete mass acquisition history 
(Gardner 2001, Vitvitska et al. 2001, Maller, Dekel \& Somerville 2002). 
In this scenario, orbital angular momentum is transferred to the remnant's
internal angular momentum during a merger. Statistically, the net
angular momentum of a large number of mergers would be zero (if infall occurs
from random directions). However, due to the low number of major 
mergers that occur during a halo lifetime, randomization is ineffective and
there remains an imprint on the halo's spin parameter by the final (and often
only) major merger. 
Mergers of dark matter halos and galaxies thus moved to the center of interest 
not only in the field of early type galaxy formation but also in the context 
of angular momentum acquisition (Burkert \& D'Onghia 2004. D'Onghia \& Burkert 2004). 

The angular momentum of dark matter halos is usually parametrised in terms 
of a dimensionless quantity, the spin parameter $\lambda$. This parameter 
 appears in two versions; the first, original one was introduced by Peebles (1980).
A revised formula was later proposed by Bullock et al. (2001) in order to 
overcome the energy dependence of the classic version. 

In no study so far the differences between the classical $\lambda$ and the 
revised version $\lambda'$ have been investigated in detail. 
This  is the main ambition of our work. For this purpose we analyse
the data of 82 dissipationless numerical simulations of dark matter halo 
formation in a comoving cube of 20 Mpc starting with redshift z=2.8. In 
order to understand our results we also investigate the connection 
between the spin parameters and the time dependent virial coefficient 
$\eta=2\ek/|\ep|$. 

In \S 2 we describe the numerical simulations in detail. 
\S3 analyses of the merger anatomy and statistics. 
A detailed investigation of the spin transfer in
merger and accretion events is given in \S4 and \S5, with special attention
drawn to the differences between the two $\lambda$-versions. In \S 6 we 
investigate the virial parameter of dark halos and it's connection with the 
spin parameters. The global time evolution of $\lambda$ and $\lambda'$ is  
shown in \S 7. Finally we summarize and discuss our results in \S8.


\section{Numerical Simulations}

We employ the {\it GRAFIC} code which is part of the {\it COSMICS} 
package developed by E. Bertschinger (Ma \& Bertschinger 1995) to set up random 
Gaussian initial conditions. 
In total, 82 dissipationless cosmological {$\Lambda$}CDM 
simulations are performed with different random seeds for the initial
fluctuation field, using a N-Body code with a Barnes-Hut tree method 
(Barnes \& Hut 1986) in combination with the special purpose hardware 
{\it GRAPE-3} (Makino \& Funato 1993, Ebisuzaki et al. 1993). 
Each run represents a 20{$Mpc$} comoving cube with the standard 
cosmological parameters  
$\Omega_m=0.7$, $\Omega_{\Lambda}=0.3$, $h=65$ and {$\sigma_8=1.13$}
\footnote{Note that these simulations were performed earlier than the current
WMAP results ($\sigma_8=0.9\pm0.1$, see Spergel et al. 2003) were published. Our value for $\sigma_8$ is following the 
expression $\sigma_8=(0.5\dots0.6)\Omega^{-0.56}$ from White, Efstathiou \& Frenk (1993). 
We don't  expect this discrepancy to affect our results. Especially results 
concerning individual merger events should be completely unaffected.}. 
The simulations start at redshifts $40<z_{init}<53$ (depending on when the 
maximum overdensity reaches unity)  and follow the nonlinear evolution of 
$64^3$ particles until $z=0$. Position and velocity data are stored at 21 
different redshifts mar\-king approximately equal time intervals.  
The gravitational potential is smoothed within 7.8 comoving kpc around 
individual particles, each of which represents $1.07\times 10^9\msol$ of 
cold dark matter. 

We identify halos using the friends-of-friends (FOF, see e.g. G\"otz, Huchra \& Brandenberger 1998) 
algorithm with a linking
length parameter of 0.2 in units of mean particle distance. Each FOF cluster
of particles is truncated at a radius $R_{\rm vir}$ which defines a 
sphere around
the most bound particle within which the mean density is 
$\Delta(z) \rho_c(z)$, $\rho_c(z)$ being the cri\-ti\-cal density at 
redshift $z$. Following the
spherical collapse model, $\Delta(z)$ would be constant in an Einstein-de 
Sitter universe, with $\Delta=18\pi^2\approx 178$ (see e.g. Peebles 1980). 
Due to the model dependence of the density parameter $\Omega(z)=(\Omega_m-1)^3
\Gamma(z;\Omega_m,\Omega_{\Lambda})$  this does not hold for a low 
density model like the one considered here. Neglecting radiative contribution
to the mean cosmological density, $\Delta(z)$ takes the general form
\begin{equation}
\Delta(z)=18\pi^2+82f(z)-39f(z)^2,
\end{equation}
where
\beq
f(z)=\frac{\Omega_m(1+z)^3}{\Omega_m(1+z)^3+\Omega_{\Lambda}}-1
\eeq
(Bryan \& Norman 1998).

Our simulations yield a total number of 10563 halos, each with more than 100
particles at $z=0$. In order to avoid a mass truncation bias in our merger 
trees, we consider only halos which contain at least 500 particles
at present epoch, corresponding to a minimum halo mass of 
$5.35\times10^{11}\msol$ which leads to a sample of 2383 halos; 
particle clusters as small as 50 particles ($1.07\times10^{11}\msol$) are 
nevertheless taken into account as progenitors and satellites at higher 
redshifts. Note that throughout this study we didn't detect any mass
dependence of our results, which therefore should be valid for lower mass
halos as well. 

We construct merger trees for the complete sample of halos with 
$M(z=0)>5.35\times10^{11}\msol$ in the following way: A major merger (MM) is 
defined to be an event where  two initially separated halos appear FOF-bound
in the next data output with 3 restrictions:
\begin{itemize}
\item the less massiv, infalling
clump (denoted as the {\it satellite}) has a fraction of at least 
$m_h\ge1/5$ of the larger clump's (the  {\it halo's}) mass.
\item the merger remnant
contains more than 60\% of the initial satellite's particle reservoir 
\item the massive progenitor halo has a mass of at least 40\% of its mass
at $z=0$. 
\end{itemize}
The first restriction  separates mergers from events which we will denote 
as "accretion" or ``minor mergers'' (mM) throughout. 
Note that we do not distinguish
between minor mergers and accretion; both terms correspond to any 
detected
matter acquisition which increases the progenitor's mass by less than 20\%.
The second criterion guarantees that the mass growth after the
preceding time step does in fact originate from the satellite infall and not
from accretion or mM events which may well happen during the same 
time period.
The third point finally rules out mergers which, due to the low mass of
the remnants, do not have a
significant impact on the dynamics of the final halo or galaxy at $z=0$. 
It is clear that our particular choice of the parameter values (60\% and 
40\% in the 2nd and 3rd item, respectively) is somewhat arbitrary and lacks
a quantitative justification. Nevertheless it seems plausible that this
choice of numbers is suitable for the purposes described above. 
By identifying all progenitor-remnant relations throughout all data
outputs we construct a list of progenitors of each halo at $z=0$ as well 
as the according mass acquisition history. 


\section{Merger Statistics and Orbital Parameter}

Before we are going to examine the spin evolution of dark matter halos in 
detail we want to  ''set the stage'' by briefly discussing the abundance 
and the orbital characteristics of major mergers. 

\subsection{Orbital Parameters}

Certainly the main quantity that characterizes the dynamics of a major merger 
is the mass ratio of the involved halos. The dynamical properties of the 
remnant, however, depend on both the internal and orbital energy 
and angular momentum contributions. If we describe the approach of two
halos following the Kepler laws for classical two body systems (which is 
well justified as long as the objects are spatially separated), the orbital 
energy and angular momentum content of the system is fully described by their 
orbital parameters. The  $\it eccentricity$ is usually defined by 
\beq
\epsilon=\left[1+\frac{2EL^2}{\mu (Gm_hm_s)^2}\right]^{1/2},
\eeq
where $E$ and $L$ are the total orbital energy and angular momentum, $\mu$
is the reduced mass, $G$ the gravitational constant and $m_h$ and $m_s$ are
the masses of the halo and the satellite, respectively. The second parameter,
the {\it pericenter distance} corresponds to the fictitious closest approach 
of the objects (pretending the system would continue to behave like a perfect 
two body system long enough) and is given by 
\beq
\rp=\frac{L^2}{\mu Gm_hm_s(1+\epsilon)}.
\eeq

We find that 90\% of all mergers have bound orbits ($\epsilon<1$) and
71\%  have values of $0.9<\epsilon<1$. This clear
maximum of $p(\epsilon)$ close to unity  can be considered as the main
feature of the eccentricity distribution. Note
that initially unbound hyperbolic orbits are as well compatible with 
merger events since orbital energy can be transformed to internal one during 
the merger.

For the pericenter distance $\rp$ we find a probability distribution which 
can be well described by $P( \rp/\rv )=A\exp(-\alpha\times\rp/\rv)$, 
where $(A,\alpha)\approx (0.089,5.0)$ for $\rp/\rv>0.2$ and
$(0.12,6.3)$ for $\rp/\rv<0.2$. The mean relative pericenter distance turns 
out to be $0.18 \rv$.

We finally checked the distribution of the angles $\beta$ between the spin 
planes
of halos and infalling satellites. In agreement with expectations we find a 
distribution that is well fitted by $dN/d\beta\propto\sin\beta$. 

Khochfar \& Burkert (2003, KB2003) investigated in very detail the statistics of various merger 
orbital parameters using a huge data sample provided by the VIRGO 
consortium. Apart from minor deviations we find the  results from this section 
to be in good agreement with KB2003. 

\subsection{Merger Statistics}

\begin{figure}
\vspace{1cm}
\centerline{\includegraphics[angle=0,scale=0.37]{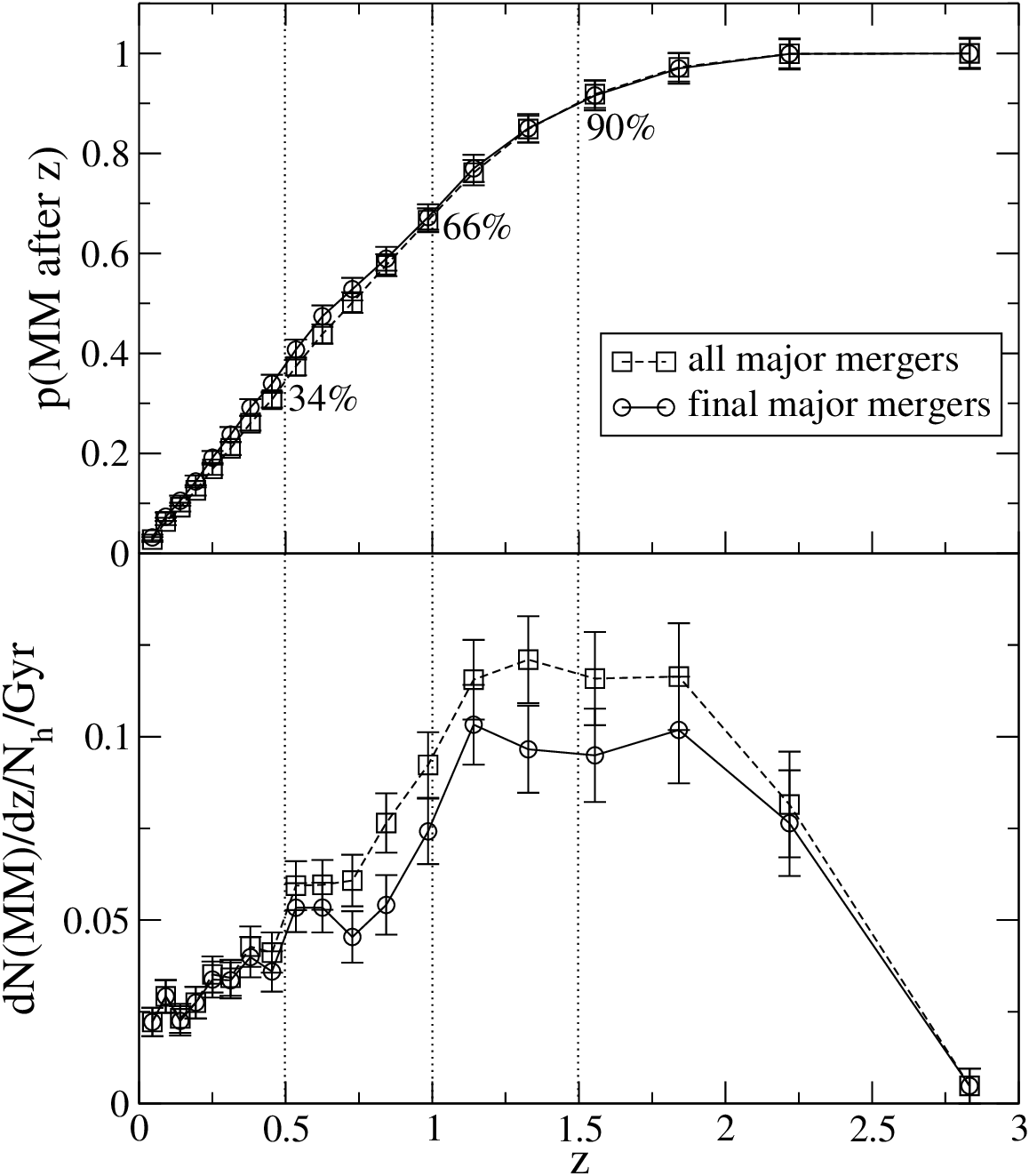}}
\caption{Top panel: cumulative redshift distribution of major mergers. 
The vertical axis shows the fraction of mergers later than redshift $z$
in our simulations. 
Lower panel: Merger rate at redshift $z$, i.e.~the number of mergers per 
unit time normalized to the total number of halos with masses above 
$5.35\times 10^{11}M_{\odot}$. \label{mmdist}} 
\end{figure}

The internal structure of dark matter halos and their embedded galaxies 
is to a high degree sensitive to their merging histories. Strong evidence 
exists now that elliptical galaxies originate from MM events 
(see Toomre \& Toomre 1972, Naab, Burkert \& Hernquist 1999, see Burkert \& Naab 2003, 2004 for a reviews).
As will be discussed in this paper, the 
evolution of a dark halo's spin parameter seems as well to be a matter of the 
mass growth history (see also D'Onghia \& Burkert 2004). Therefore it is worth to illustrate the mean merger 
history of our sample of dark halos throughout their evolution.
The top panel of Fig.\ref{mmdist} 
shows the probability for a dark halo to 
have a (final) MM at redshift smaller than $z$ (if it has a MM at all), 
while the lower one shows the number
of (final) MMs per time, normalized to the total number of
halos with masses higher than $m_{min}>5.35\times 10^{11}M_{\odot}$ 
(corresponding to 500 particles 
minimum). The so defined merging rate shows a broad maximum  between
redshift 1.2 and 1.8, where the rate is roughly constant. After $z\approx 1$
as well as before $z\approx2$ the rate is continuously decreasing to a value of 
$\approx 0.02\pm 0.004/N_h/Gyr$ and $0.05\pm0.005/N_h/Gyr$ at $z=0$ and $z=2.8$,
respectively, 
where $N_h$ is the total number of halos above $m_{min}$ at the 
corresponding redshift. Roughly one third of all MMs occur in redshift 
intervalls of  $z=0-0.5, 0.5-1$ and $1-\infty$ (see top panel of 
Fig.\ref{mmdist}).
This shows  that the spin parameter of dark halos may undergo changes 
throughout it's whole lifetime, if the spin is affected by these merger 
events. 
\vspace{.7cm}

\section{Spin Transfer in Accretion and Merger Events}\label{parstrans}

In the commonly established scenario, cosmic structures acquire angular 
momentum by tidal interaction with the surrounding matter distribution 
(Barnes \& Efstathiou 1987, Padmanabhan 1993 and first Hoyle 1949). This mechanism works 
most effectively during the early expansion and collapse phase, when the large spatial 
extension of the protohalo offers a large  lever for tidal forces. The 
Zel'dovich approximation is a powerful tool for an analytical 
treatment of the spin acquirement and it turns out that angular momentum 
is growing linearly with time until the point of turnaround is reached. 

More recently, cosmologists became aware of merger events to have a 
significant impact on the halos' spin by the transfer of orbital to 
internal angular momentum  (Gardner 2001, Vitvitska et al. 2001 (VK01)). It 
appears plausible that the quantitative evolution of spin transfer 
in mergers is depending on the corresponding orbital parameters.
This complicates the investigation of the spin evolution in so far as
we have to deal with nonlinear physics  which rules out an ab initio
analytical description like the linear tidal torque theory. 

\subsection{Spin Transfer Statistics}

To explore the merger impact on the halo spin parameter in detail, the most 
basic investigation is to analyse the probability
of a single merger event to rise (or decrease) the spin of a halo by a 
certain amount. The alteration of spin is quantified simply by the spin 
transfer
\beq
\tl:=\lf/\li, 
\eeq
where $\lf$ is the (final) spin parameter of the
merger remnant and the initial  $\li$ corresponds to its main progenitor. 
We want to highlight the characteristics and differences of both
established versions of dimensionless spin parameter definitions throughout 
this paper. The earlier classical version introduced by Peebles (1980) reads
\beq\label{sparam1}
\lambda=\frac{J\sqrt{E}}{GM_{\rm vir}^{5/2}}.
\eeq
Here $J$ is the modulus of the total angular momentum, $E$ is the absolute 
value of the total (kinetic plus potential) energy and $M$ is the mass of the 
halo under consideration. All quantities are calculated using the set of 
particles within the  virial radius $\rv$. 
Bullock et al. (2001) later defined an alternative version, namely 
\beq\label{sparam2}
\lambda'=\frac{J}{\sqrt{2}M_{\rm vir}R_{\rm vir}V_{\rm vir}},
\eeq
where $V_{\rm vir}=GM_{\rm vir}R_{\rm vir}^{-1}$ is the circular velocity at 
the virial radius $R_{\rm vir}$. This 
latter (dashed) quantity is more frequently applied in practical analysis
due to its independence of the density profile. 
Both quantities are nevertheless approximately identical for halos with 
NFW profiles (Navarro, Frenk \& White 1996) and  concentration parameter $c\approx 10$ 
as well as isothermal halos truncated at $R_{vir}$ (Maller, Dekel \& Somerville 2002).

Fig. 2 shows the $\lambda$-distribution for all halos at $z=0$
and for the merger and accretion dominated subsamples of halos as well. 
One can clearly see a shift between each sample corresponding to the according
spin transfer behaviour: As to expect, MM dominated halos are shifted to higher 
$\lambda$ values, mM halos occupy lower values.  In each case the distribution is 
roughly lognormal. Here we do not show the distribution of $\lambda'$ 
but note that we find a similar behaviour.

\begin{figure}
\vspace{1cm}
\centerline{\includegraphics[angle=0,scale=0.37]{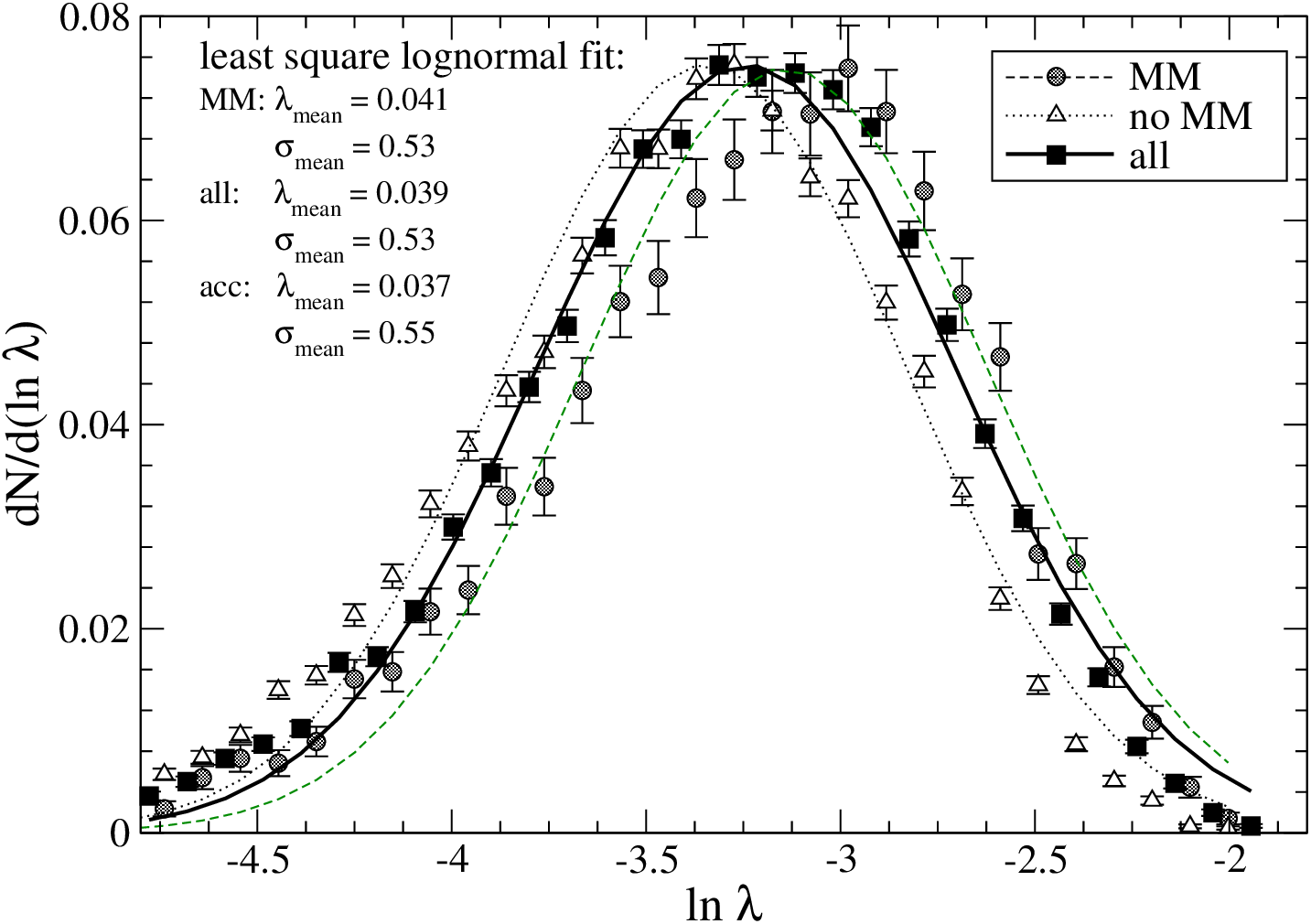}}
\caption{Distribution of the (classical) spin parameter $\lambda$. Each sample (halos
with MMs, without MMs and full sample) is following roughly a lognormal 
distribution (see e.g. Cole \& Lacey, 96). Depending on the merger history the
distribution is shifted to higher (MM sample) or lower (mM sample) 
values.} 
\end{figure}

\begin{figure}
\vspace{0.7cm}
\centerline{\includegraphics[angle=0,scale=0.37]{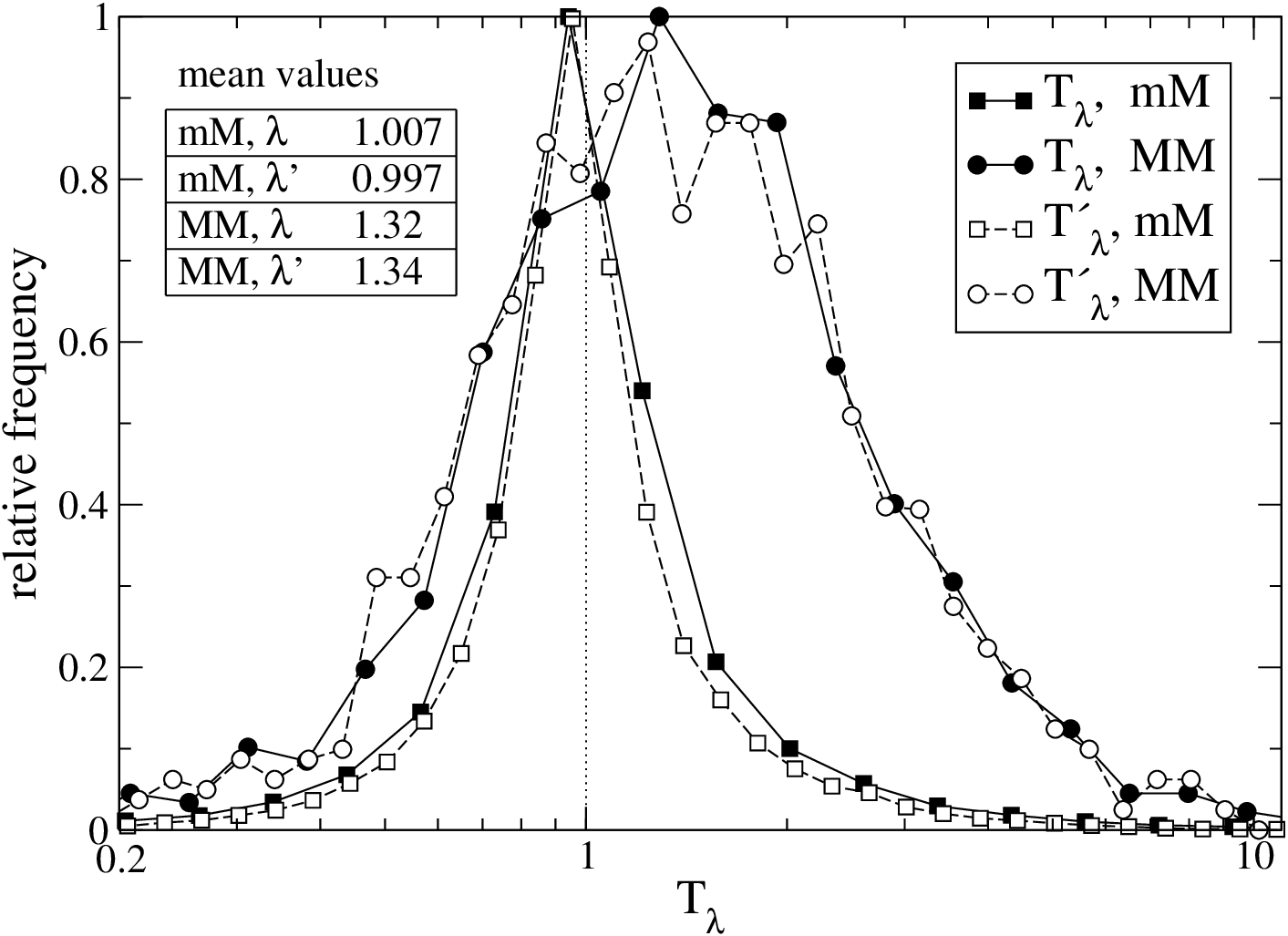}}
\caption{Distribution of the spin transfer rates $\tl=\lf/\li$ (full lines) and 
$\tl'=\lf'/\li'$ (dashed lines)  for major 
(circles) and minor mergers (squares). Clearly major mergers 
provide a source of angular momentum: roughly two third of all MMs increase 
the spin parameter; in the case of minor mergers less than half of all events 
increase $\lambda$ by a moderate amount. Note that the behaviours of $\tl$ 
and $\tl'$ are hardly distinct for either merger type. See the discussion in 
the text for details. \label{sratiodist} }
\end{figure}

Fig.\ref{sratiodist} 
shows the distribution of the spin transfer rates 
$\tl$ and $\tl':=\lf'/\li'$ for the case of major ($m_s>0.2m_h$) and minor 
merger/accretion ($m_s<0.2m_h$) events. In the following discussion all 
numbers concerning $\lambda'$ are listed in brackets behind the  
numbers corresponding to the undashed $\lambda$.

Fig.\ref{sratiodist} clearly shows that MMs rise the spin parameter by a 
significant amount: 68\% (67\%) of all 
MMs rise $\lp$. In more than 50\% of all MMs, $\lp$ is
increased by a factor of 1.30 (1.31). This is in good agreement with the 
previous study of VK01 who found a  median of 1.25 for the dashed quantity 
$\tl'$. For the according mean values we find $\oli\tl^{(\prime)}$ are 
1.32 (1.34). 

As expected the situation is different for the case of mM and accretion 
events. Again there are both, spin-rising and decreasing events, but now with 
a clear shift to a lower $\tl$ regime: The  distribution of $\tl$ in 
this case is significantly narrower and approximately symmetric around unity, 
with a linear median value of $\tl=0.98 (0.99)$ and a mean of 1.007 (0.997). 
53\% (56\%) of all events decrease the spin parameter. 

It turnes out that the behaviour of both spin parameters is very similar in
the above context, regardless of the density profile dependence of 
$\lambda/\lambda'$ which should lead to visible differences especially
during the merging phase, where the density distribution is strongly time
dependent.
Nevertheless the tiny difference of the $\oli\tl$ and $\oli\tl'$ distribution 
in the 
accretion case has striking effects on the global time evolution of 
$\lambda$ and $\lambda'$. We will discuss this important issue in detail 
in \S \ref{timeevol}. 

\subsection{Dependence on Orbital Parameters}

\begin{figure}
\vspace{1cm}
\centerline{\includegraphics[angle=0,scale=0.37]{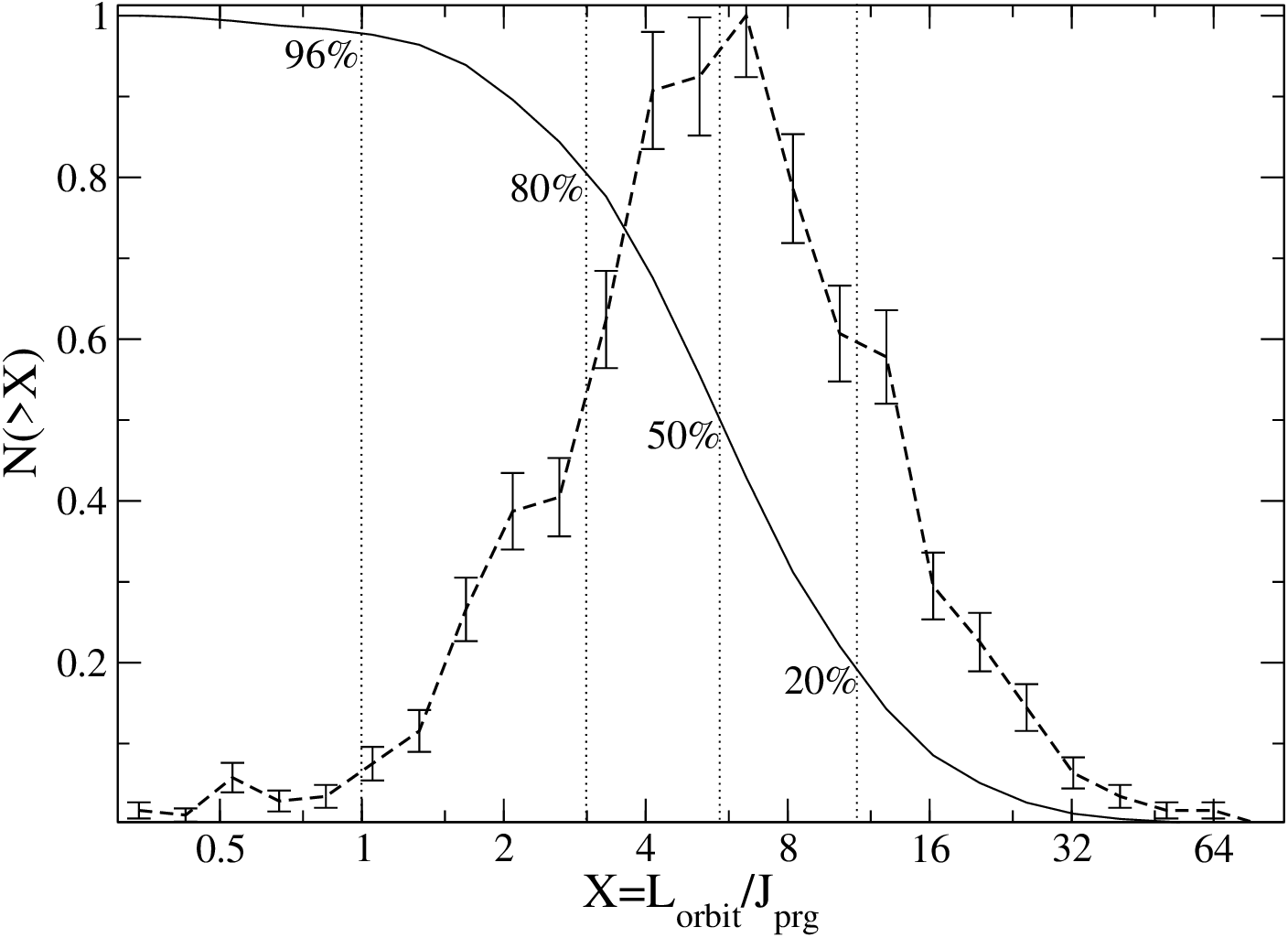}}
\caption{Cumulative (full line) and differential (dashed line) 
distribution of the ratios of orbital to internal angular momentum.
In 50\% of all mergers $L_{orbit}>6\times J_{\rm prg}$. In less than
4\% of all cases $J_{\rm prg}>L_{orbit}$. The
diagram demonstrates that the spin transfer is dominated by the
orbital angular momentum. \label{lversuss} } 

\end{figure}

\begin{figure}
\vspace{1cm}
\centerline{\includegraphics[angle=0,scale=0.37]{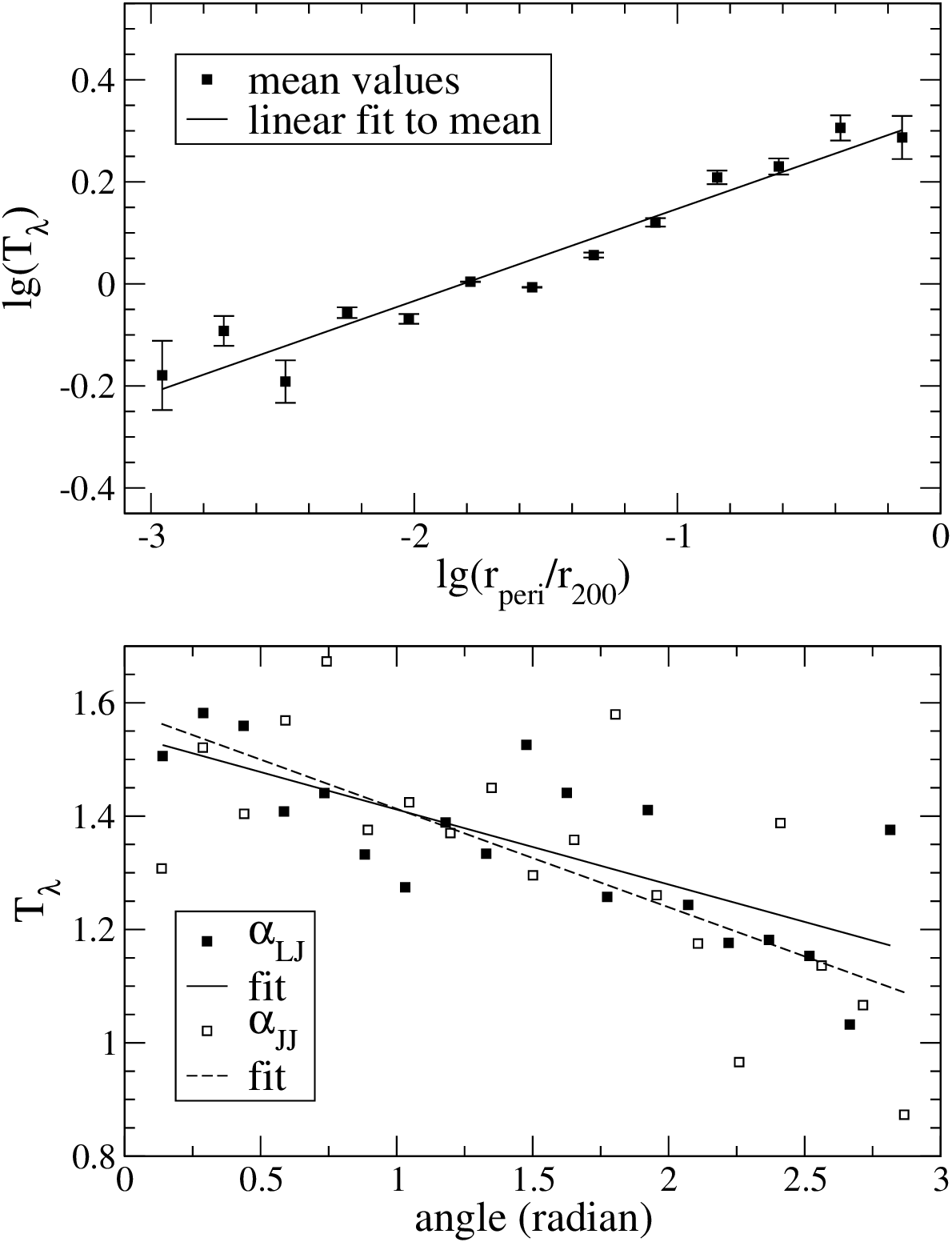}}
\caption{Top: Spin transfer $\tl$ vs.~relative pericenter distance. The
diagram shows a tight correlation following
$\tl\propto(\rp/\rv)^{0.18}$. The error bars reflect the Gaussian tolerance for
the mean values rather than the standard deviation of the individual merger 
data. Bottom: $\tl$ vs.~ spin-spin and 
spin-orbital angular momentum angle ($\alpha_{\vec L\vec J}$ and 
$\alpha_{\vec J \vec J}$ respectively). As typically
$L_{orbit}/J_{\rm prg} > 1$ the $\tl-$angle correlation is
clearly weaker than the relation between  $\tl$ and $\rp$ \label{sratioparams} }
\end{figure}

As mentioned before, the plausible explanation for the qualitative
behavior just described is that orbital angular momentum and the spin of
the satellite are transferred
to internal spin of the remnant during the merging process, according to
\beq\label{lsum}
\vec J_{\rm remn}\approx\vec J_{\rm prg}+\vec J_{\rm sat}+\vec L_{\rm orbit}
\eeq
where the indices 
represent remnant, progenitor, satellite and the system's orbit 
respectively. Exact equality is unlikely since some dark matter particles 
are lost during the merging process.  $\tlp$ is more or less 
sensitive to the re\-la\-tive orientation of the vectors on the right hand 
side of Equ.~(\ref{lsum}), depending on the relation of their absolute values.
Since we are discussing major merger events it is justified to assume 
satellites and halos to have internal angular momenta of the same order 
of magnitude. 
From Fig.\ref{lversuss} one can see that only in a small fraction (4\%)
of all MMs the halo internal angular momentum is larger than 
$L_{\rm orbit}$. 
In 80\% of all events $L_{\rm orbit}>3\times J_{\rm prg}$  and still for
every third case $L_{\rm orbit}>7.7\times J_{\rm prg}$. This demonstrates 
directly that the spin transfer is dominated by the orbital angular momentum. 
We find a small but detectable dependence of  $\tl$  on the vector 
angles $\alpha_{\vec L\vec J_{\rm prg}}$ and 
$\alpha_{\vec J_{\rm prg} \vec J_{\rm sat}}$. The bottom panel of 
Fig.\ref{sratioparams} 
shows a linear fit
through the median $\tl-$angle plot. Though suffering from a large 
scatter, there is a clearly visible trend of increasing spin transfer with 
decreasing spin-orbit- and spin-spin-angle.

If the transfer of angular momentum is dominated by 
$L_{\rm orbit}$ we should also expect a more elementary dependence of
$\tl$ on the pericenter distance since (for fixed eccentricity 
$\epsilon$) $\rp$ is proportional to the satellites' impact parameter 
$p_{\rm i}$ which, for its part, satisfies $L_{orbit}\propto vp_{\rm i}$.
 In Fig.\ref{sratioparams} 
(upper panel)  $\tl$ is plotted against 
$R_{peri}/R_{200}$ which is well fitted by the power law 
\beq
\tl\propto \left[ \frac{\rp}{R_{200}} \right]^{0.18}
\eeq
over a $\rp/R_{200}$ range of almost 3 orders of magnitude. This clearly 
reflects the leading role of the impact parameter for the amount of spin transfer  
in a merger event. We also investigated the according behaviour of the dashed
spin transfer $\tl'$ and didn't find any significant difference. 

\section{Before, During and After Merging --- Rising and Declining Spin}
\label{sec_before}

\begin{figure}
\vspace{1cm}
\centerline{\includegraphics[angle=0,scale=0.37]{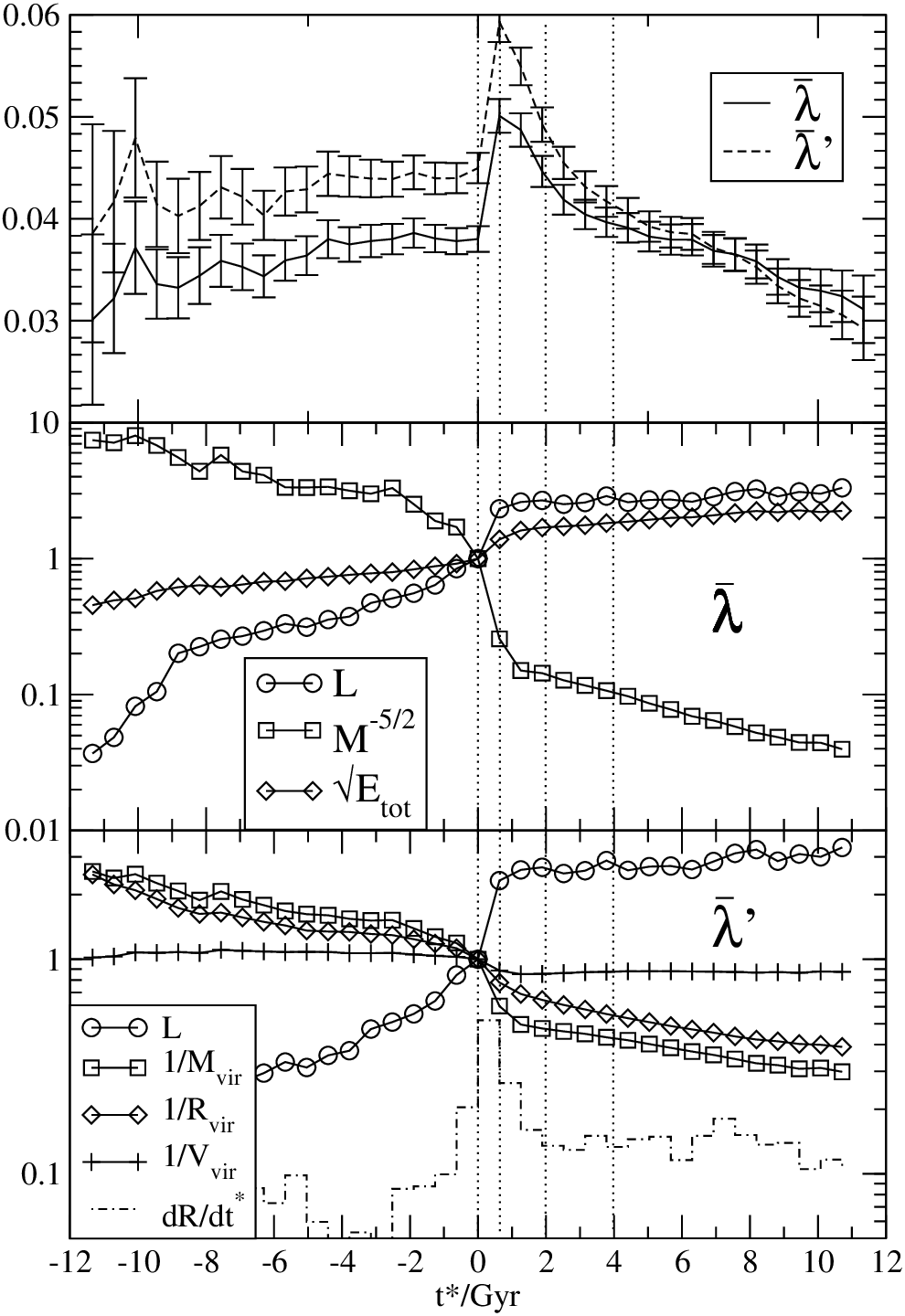}}
\caption{Top panel: Evolution of the mean spin parameters $\oli\lambda$ and 
$\oli\lambda'$.  The time axis $t^*$ denotes the time for 
each individual halo relative 
to its final major merger event at $t^*=0$. The sudden rise of $\oli\lp$ 
right after the merger onset is followed by a less rapid decline.
The middle and bottom panel show the (relative) time evolution of the 
quantities that make up the spin parameters (see Eqs.~({\ref{sparam1}) and 
(\ref{sparam2}})). In both cases, the sudden rise 
of the angular momentum (open circles) is responsible for the fast increase of the
spin parameter after $t^*=0$. Note also the dotted-dashed line in the bottom
panel, which illustrates the differential growth of the virial radius. 
See text for further discussion.  \label{sparevolmm} }
\end{figure}

Major mergers at first appear to be the main sources for rising spin 
parameters in the nonlinear evolution of dark halos. The upper panel of 
Fig.\ref{sparevolmm} however offers a more complex view: it shows the 
evolution of the mean $\overline\lambda$ along the ``fictitous'' time axis 
$t^*$, where the latter 
variable represents an individual time scale for each halo with the zero 
point $t^*=0$ shifted to the halo's final major merger event. 
In other words, the individual curve 
$\lambda_j(t)$ of each halo $j$ is shifted along the (real) time axis such 
that its final MM  happens at the same fictitious time $t^*=0$. 
The average is taken within time intervalls of 0.6 Gyrs.

For $t^*<0$ the mean spin parameter is almost constant or at most slightly 
increasing with a rate of 
$d\overline\lambda/dt \approx (4.8\pm0.8)\times 10^{-4}{\rm Gyr}^{-1}$. 
Within the first 0.6 Gyrs after the merger onset 
$\overline\lambda$ rises fast by 30\%, followed by a 5 Gyrs lasting 1st 
phase during which the spin decreases again with a power law 
$\overline\lambda(t^*)\propto 1/(t^*)^{0.15}$, and a 2nd phase with a further 
decline, following a quadratic function. 

The $\lambda'(t^*)$ curve looks similar. There is a significant 
peak after the merger onset and a modest decline later on. 
Note however that both spin parameters have equal mean values only at late 
fictitous times $t^*$, i.e.~well after the final mergers. For times $t^*<3$
Gyr $\lambda$ is significantly lower than $\lambda'$, and the difference
increases towards earlier times. For $t^*<0$ $\lambda'$ is also roughly 
constant. 

To understand the behaviour of the mean spin parameters $\oli\lp$ in further 
detail, 
it is instructive to explore the time evolution of the quantities which 
constitute the spin parameters, namely the angular 
momentum $L(t^*)$, the total energy $E(t^*)$ and the total mass $M(t^*)$ for 
the case of $\lambda$ (Equ.~(\ref{sparam1})) and the virial radius
$R_{\rm vir}$ and velocity $V_{\rm vir}$ for $\lambda'$
(Equ.~\ref{sparam2}). The results are shown in the middle and lower panel 
of Fig.\ref{sparevolmm}.
Each curve is normalized to unity at the merger 
onset $t^*=0$.

We consider the middle panel of Fig.\ref{sparevolmm} 
first, representing the undashed mean $\oli\lambda$. 
Already for $t^*<0$ the permanent smooth matter infall  
causes a steady growth of the total energy absolute value $|E|$, angular 
momentum $L$ and mass $M$ within the redshift dependent virial 
radius. Until the merger's onset the spin rising effect of $\sqrt E$ and $L$
compensates and somewhat exceeds
the decreasing effect of the $M^{-5/2}$ term in the 
denominator of Equ.\ref{sparam1}. 
While the values of  $L$ and $\sqrt{E}$ rise by factors of 
approximately 5 (neglecting the initial rise which is likely due to the poor
statistics for $t^*<-8$ Gyrs) and 2 respectively, $M^{-5/2}$ drops by a factor 
of $\approx 7$ during the same period.

At $t^*=0$, a large satellite enters the halo and the mean angular momentum 
and the total energy rise by factors of 2.3 and 1.4 respectively within only
0.6 Gyrs, while the total mass term ($M^{-5/2}$) is decreasing by a factor of 
$\approx 0.25$. 
One must be carefull not to simply multiply these quantities in order to predict
the behaviour of $\oli\lambda$. Note that we are considering mean values 
($\overline L, \overline {M^{-5/2}}, \overline {E^{1/2}}$) here, i.e.~the 
sums of single values, while the mean spin parameter $\overline\lambda(t^*)$ 
in Fig.\ref{sparevolmm} is a sum of the products of individual 
$\sqrt{E_j},L_j$ and $M_j^{-2.5}$ values. Both quantities, 
$\overline\lambda(t^*)$ and 
$\overline L(t)\times\overline {E^{1/2}}(t^*)\times \overline {M^{-2.5}}(t^*)$ 
would be comparable only for small variances of the single values.

We can nevertheless draw some qualitative conclusions on the roles of $L, E$ 
and $M$ for the evolution of the spin parameter $\ovl(t^*)$.  Obviously, 
the rise of angular momentum is the quantity mainly responsible for the 
sudden rise of $\ovl$ after $t^*=0$ while the total energy seems to 
contribute to a minor 
degree. After $t^*\approx 0.6$ Gyr, when $\ovl$ is decreasing, the figure
shows that this is due to the change of the mean mass, which is 
continuing to increase by a factor of $\approx 5$ due to accretion between 
the final merger and $z=0$ while  $\overline L$ and 
$\overline{E^{1/2}}$ are not changing significantly during the same period. 

The lower panel of Fig.(\ref{sparevolmm}) shows the situation for the 
case of $\lambda'$: Before $t^*=0$ the increasing mean mass and virial 
radius is almost canceling the effect of the increasing angular momentum. 
At $t^*=0$ the spin parameter is rising by more than 30\% within 0.6 Gyrs, 
caused by the fast rising angular momentum as in the (undashed)
$\lambda$ case. Later on, i.e.~after $t^*=0.6$ Gyr the growing mass and
radius in combination with an almost constant mean virial velocity and angular momentum
of the halo sample is responsible for $\oli\lambda'$ to decrease. 

\section{The Virial Coefficient $\eta$} 

We store our simulation data at 21 particular times 
between $z\approx 2.5$ and $z=0$ with approximately equal time intervalls 
$\Delta t\approx 6.3\times10^8 {\rm yrs}$. We therefore can detect a MM 
in a phase within a period $\Delta t$ after the individual virial radii 
of both merger partners 
had first contact. In such an early stage of merging the merger remnant is
very disturbed and
far from dynamical equilibrium. This might cause the temporary peak of the
spin parameter values at $t^* \approx 0$. To investigate this
question and quantify the relaxation of a 
halo we employ the virial parameter
\beq
\eta:={2\ek\over|\ep|}
\eeq
where $\ek$ and $\ep$ are the total kinetic and potential energies of
all particles within $\rv$. According to the virial theorem $\eta$
should approach unity for an isolated relaxed object. Note that infalling
material is affecting the surface pressure $S_p=-\int p\hat r\cdot\vec{dS}$ 
of dark halos in cosmological
simulations ($p$ is the pressure with radial direction $\hat r$ on the 
surface element $\vec dS$). 
Since the scalar virial theorem in absence of magnetic fields
reads $2\ek+\ep+S_p=0$, a momentum flux induced negative surface pressure
causes $2\ek/|\ep|>1$ (Shapiro et al. 2004), which we indeed find for the mean virial 
coefficient $\oeta$ at any redshift (Fig. 8).  Figure 7 shows the distribution of
$\eta$ and its dependence on redshift. Only a small fraction of halos have
$\eta = 1$. In addition, $\eta$ on average increases with increasing z while
the width of the distribution remains roughly constant. 
\vspace{.7cm}
\begin{figure}
\vspace{1cm}
\centerline{\includegraphics[angle=0,scale=0.37]{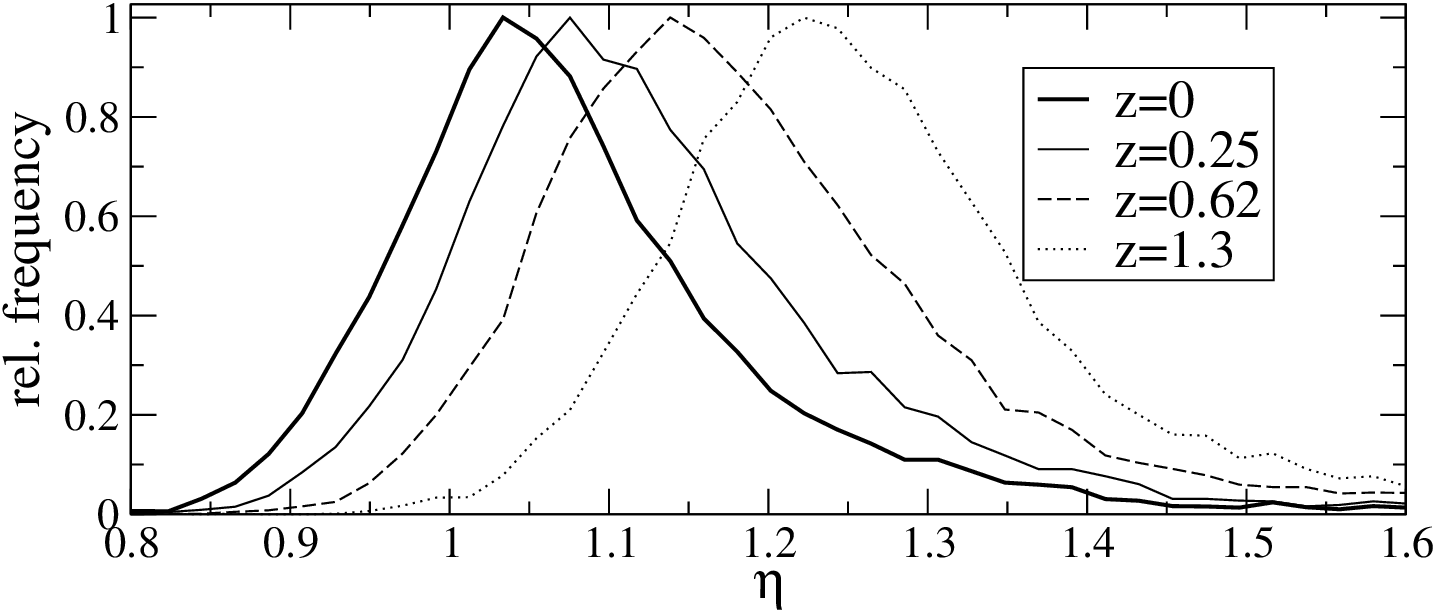}}
\caption{Distribution of the virial coefficient $\eta$ at various redshifts. Note
that even at $z=0$, the distribution doesn't peak at $\eta=1$, but is shifted
to slightly higher values due to infalling material. \label{cvirdist} } 
\end{figure}

\subsection{Global Time Evolution}

\begin{figure}
\vspace{1cm}
\centerline{\includegraphics[angle=0,scale=0.37]{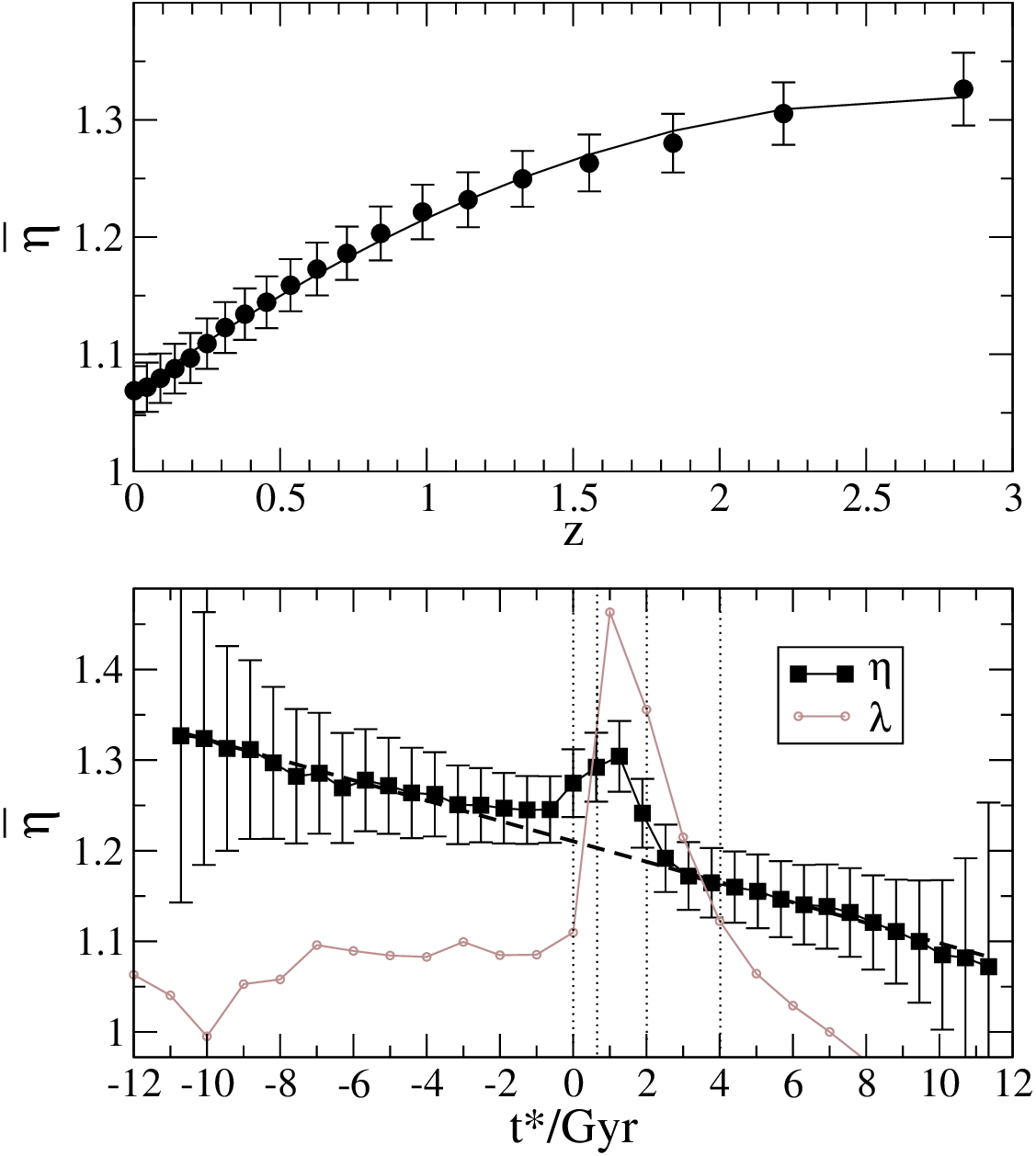}}
\caption{Top panel: Redshift evolution of the mean virial coefficient. $\oeta$ is 
decreasing with declining redshift, reflecting the tendency of halos to
become more relaxed and infall to be less important.
The quadratic decline with decreasing redshift
corresponds approximately to a linear decline with time. Lower panel: 
The linear behaviour of $\oeta$ with victitious time $t^*$ is disturbed only
by the merger event, analogous to the behaviour of the spin parameter
(grey line). Before and after the merger $\oeta(t^*)$ is described by
the same linear function (dashed line, see text for details). \label{cvirmm} }
\end{figure}

Before we are going to consider the correlation between the spin parameters
and the virial coefficient, we want to examine the time and redshift evolution 
of $\eta$. As shown in the top panel of Fig.\ref{cvirmm} the mean virial 
coefficient is clearly decreasing with decreasing redshift, well described by 
the quadratic function $\oeta(z)=-3.3\times 10^{-2}(z+2.7)^2+1.3$; this 
corresponds in good approximation to a linear decline of $\oeta$ with time, 
namely $\oeta(t)\approx 1.35-1.7\times10^{-2}t/{\rm Gyr}$. It reflects the 
general tendency of halos to approach
a more dynamically relaxed state, corresponding to $\eta\approx1$, according to the 
fact that interactions between halos in later times are less frequent and
disturb halos less than at high $z$. \\
In addition, the surface term $S_p$ becomes less important. The 
lower panel of Fig.\ref{cvirmm} shows the evolution of 
$\oeta$ with fictitous time $t^*$ relative to
the final merger (analogous to the spin parameter evolution in 
\S\ref{sec_before}). We find again the peak-like feature at $t^*=0$, similar to 
$\lp(t^*)$. At the epoch of the major merger happens, the general decline stops and 
$\oeta$ increases temporarily instead. Compared to the merger induced rise of 
$\lambda(t^*)$, the change in $\oeta$ is
measurable already before the merger onset:
More than 2 Gyrs before $t^*=0$, $\oeta(t^*)$ starts to deviate significantly 
from the linear decline which is shown by the dashed
curve in Fig. 8 and already 0.6 Gyrs before merging $\oeta$ 
experiences its fastest rise. Obviously the internal dynamical equilibrium 
of the halo is already disturbed by the proximity of a satellite; other than in
the spin parameter case, the effect on $\eta$ is preceding the
physical contact. $\oeta$ reaches it's maximum value at the same time as $\lambda$.
About 2.5 Gyrs after the merger, 
$\oeta$ is continuing to evolve as if nothing happened, i.e. $\oeta(t^*)$ 
is following the same linear behaviour as before (see Fig.\ref{cvirmm}). 
Obviously the temporary fast increase in the spin parameters shortly after a major
merger is at least partly caused by the fact that the halos are out of
virial equilibrium and the spin parameters achieve again a more typical value after
a relaxation time which corresponds to a few dynamical timescales.

\vspace{.4cm}

\subsection{Correlation between $\eta$ and $\lp$}

Peebles' classical definition
of the spin parameter (Equ.\ref{sparam1}) makes use of both the kinetic
and potential energy. Therefore one can expect a direct
correlation between $\lambda$ and $\eta=2\ek/|\ep|$, disturbed only by the 
scatter of the independent angular momentum $J$ in the definition of the spin 
parameter. The Bullock et al.~alternative $\lambda'$ is not using mechanical
energies explicitly. It should nevertheless depend on relaxation
since the virial radius which defines the region inside which $\lambda$ is
determined, is not well defined in a disturbed, unrelaxed halo.

\begin{figure}
\vspace{1cm}
\centerline{\includegraphics[angle=0,scale=0.37]{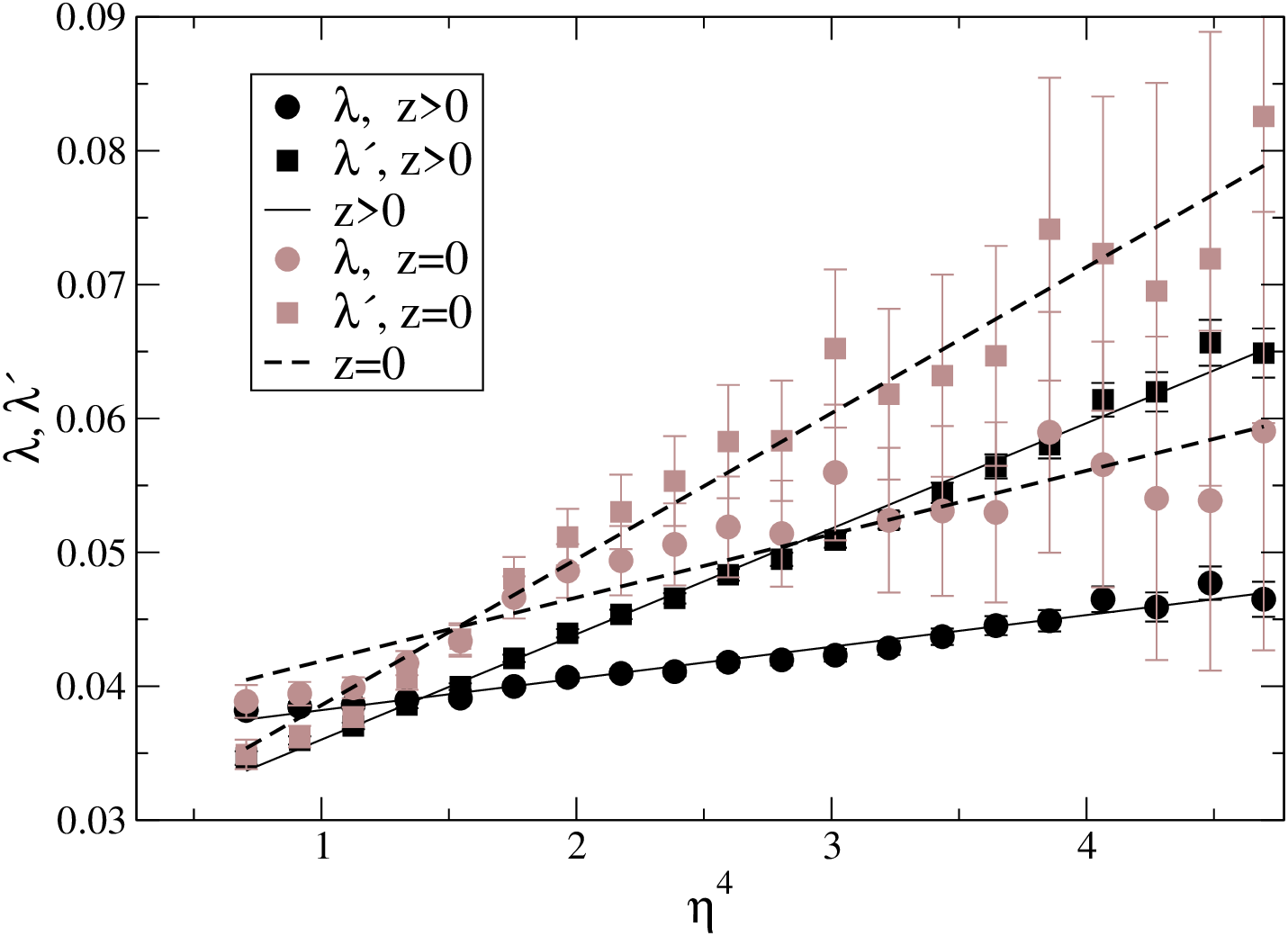}}
\caption{ Spin parameter dependence of the virial coefficient $\eta$. Each case can
be well fitted by $\lp=\alpha+\beta\times 10^{-3}\eta^4$. The fit parameter 
values are 
$(\alpha,\beta)
=(0.036,2.4)$ for $\lambda;z>0$,
$(0.028,7.9)$ for $\lambda';z>0$,
$(0.037,4.7)$ for $\lambda;z=0$ and
$(0.028,11)$ for $\lambda';z=0$.
 $\lp(\eta=0)$ is almost independent of the redshift. 
Note that ``$z>0$'' refers to the complete sample of halos excluding only
those at $z=0$. Due to the good number statistics, the 
error bars of the $z>0$ data are smaller than the symbol 
size. \label{sparcvir}} 
\end{figure}

Fig.\ref{sparcvir} shows the correlation between $\lp$ and the virial 
coefficient $\eta$. Both $\lambda$ and $\lambda'$ are increasing 
for higher values of $\eta$, following $$\lp\approx \alpha+\beta\eta^4$$
within a range of $0.9<\eta<1.5$, which contains more than 90\% of all halos.
Two clear trends are visible: 
\begin{enumerate}
\item The Bullock spin parameter $\lambda'$ is significantly stronger 
correlated with the virial coefficient $\eta$. 
\item The correlation becomes slightly weaker at higher redshifts. 
\end{enumerate}


The values of the parameters $\alpha$ and $\beta$ can be found in the caption
of Fig.\ref{sparcvir}. 

\begin{figure}[h!]
\vspace{0.7cm}
\centerline{\includegraphics[angle=0,scale=0.37]{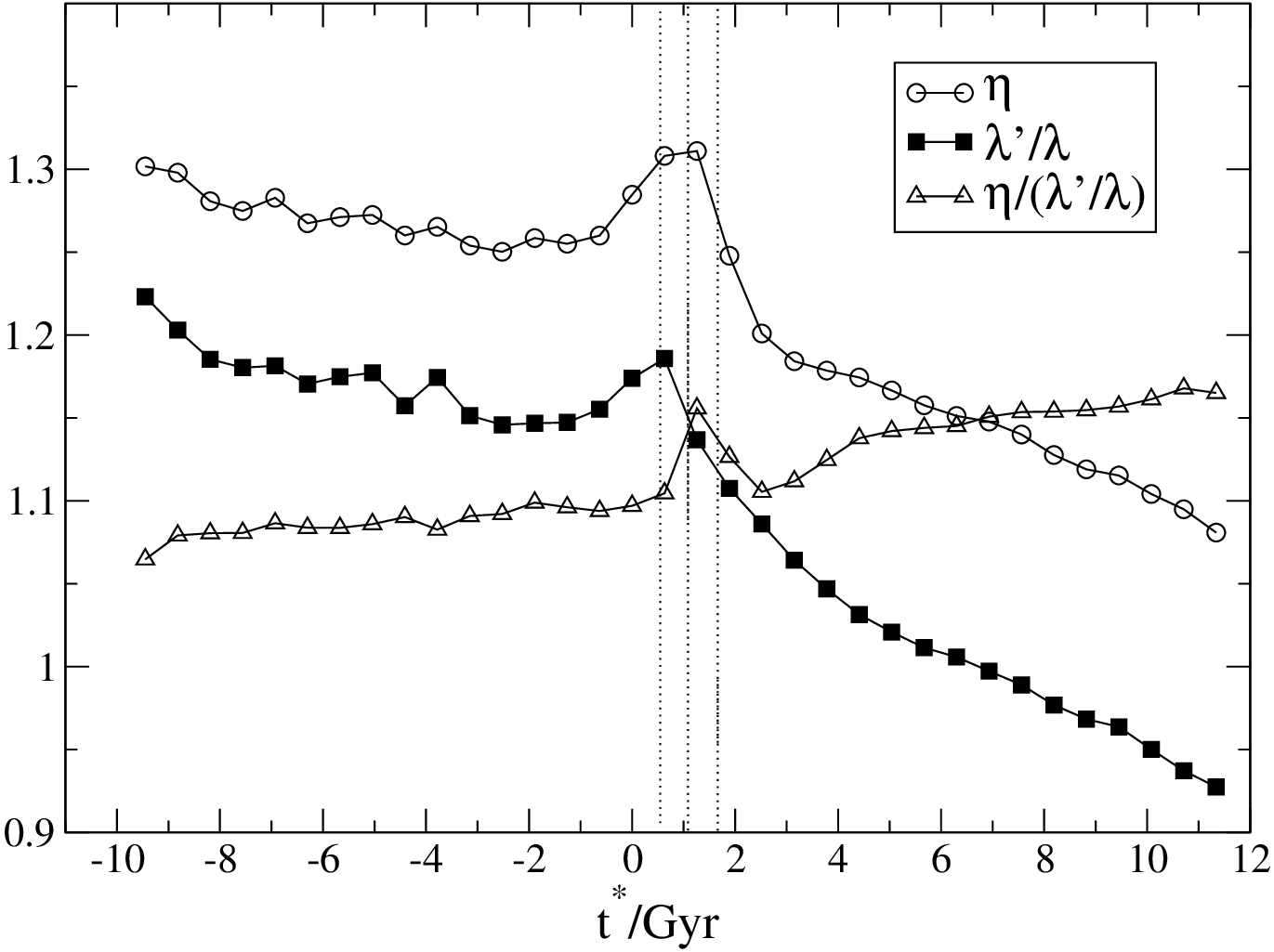}}
\caption{
Evolution of the spin parameter ratio $\lovl$ (full squares) with respect to 
the final merger event ($t^*=0$). Plotted are also the virial coefficient 
$\eta$ (open circles) and the ratio $\eta/(\lovl)$ to illustrate the strong
correlation between $\eta$ and $\lovl$, which is only disturbed for a short
time period after the major merger. \label{cvirsparratio} } 
\end{figure}

We finally examine the ratio of the spin parameters 
$\lovl$ in order to further illuminate the differences in their 
behaviours. Fig.\ref{cvirsparratio} shows that $\lovl$ is decreasing with time
-- in quite a similar manner as the virial coefficient -- again featuring 
a hump in the vicinity of $t^*=0$. The strong correlation between $\lovl$ and
$\eta$ is illustrated by the $\eta/(\lovl)$ plot in the same figure: the
linear behaviour of this quantity is only disturbed for a relatively short 
period directly after the merger onset, indicating that $\lovl$ is falling 
back to a linear evolution more quickly than the virial coefficient. Note 
that $\eta/(\lovl)$ in Fig.~\ref{cvirsparratio} is not yet affected at $t^*=0$ 
when the individual quantities already reach values close to their maximum 
ones. Thus the 
ratio $\lovl$ is stronger correlated with the virial coefficient $\eta$ 
than each spin parameter separately. This result turns out to be 
important for the next chapter.
\vspace{0.7cm}

\section{Time Evolution of the Spin Parameters}\label{timeevol}

In this chapter we want to address the question, how the sum of merging 
and accretion affects the global evolution of the spin parameters with 
cosmological time. It turned out in this work and previous ones 
(e.g. Vitvitska et al. 2001), that 
individual MM and mM events can either rise or decrease the spin parameter: 
While mMs give $\tl\approx0$ {\it on average}, it is obvious that MMs tend 
to clearly rise $\lambda$. 

The spin history of any
single halo is highly chaotic as soon as the smooth phase of spin-up increase 
by tidal torques is left behind and non linear matter acquisition comes into 
play (Peirani, Mohayaee \& de Freitas Pacheco 2003, Primack 2004). 
But if we instead consider the mean value of all halos we may still expect 
a smooth redshift evolution of the mean spin parameter due to the cumulative increase
of the total number of mergers during the cosmic evolution shown in 
Fig.\ref{mmdist}. 

\begin{figure}[h!]
\vspace{0.7cm}
\centerline{\includegraphics[angle=0,scale=0.37]{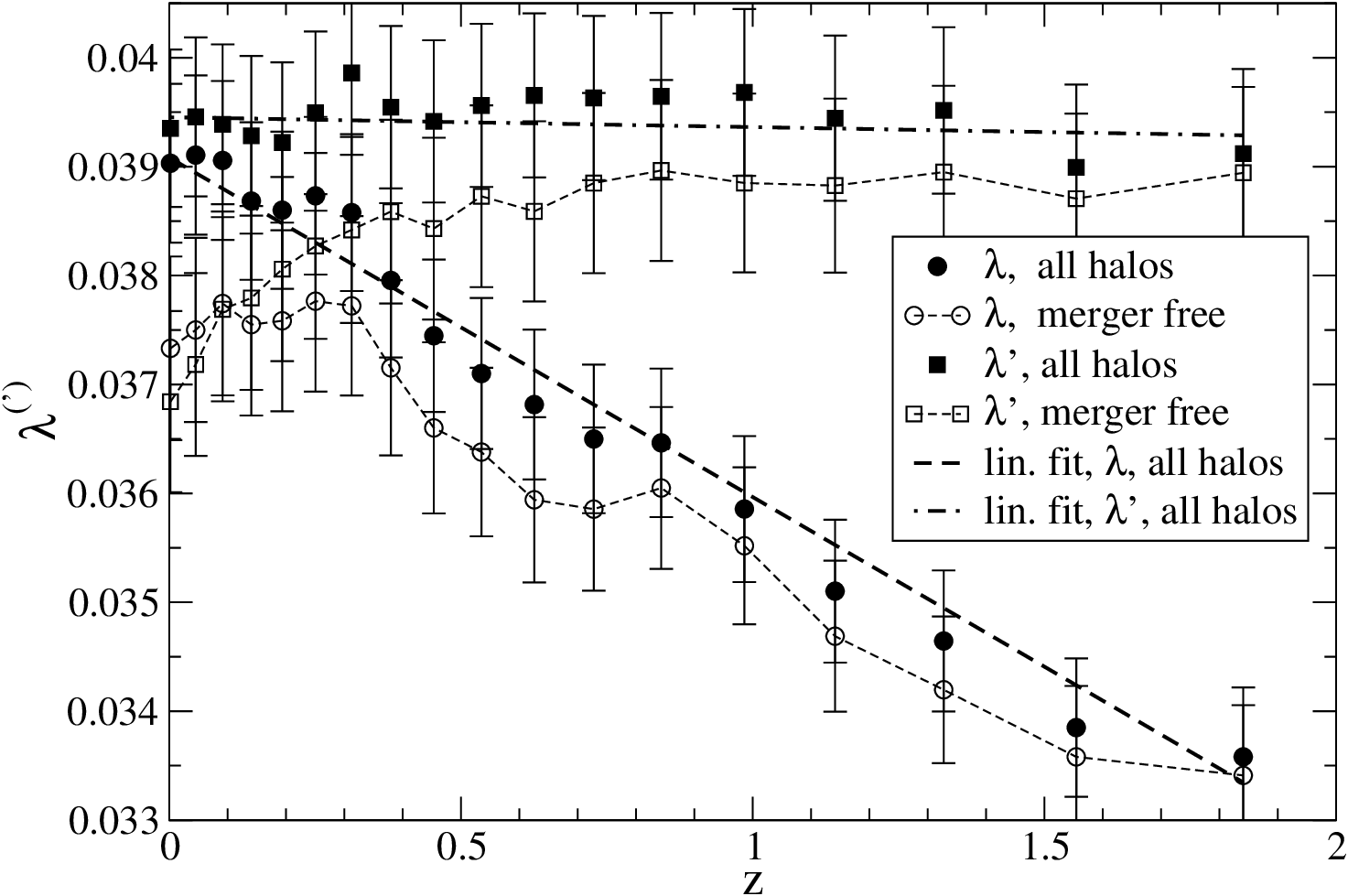}}
\caption{Redshift evolution of the mean spin parameters $\overline\lambda$ and $
\oli\lambda'$ for the mM- and the full sample of halos. 
While $\oli\lambda$ is increasing with time for both halo samples, we find  
$\oli\lambda'\approx\rm const$ for the full halo sample and 
$\oli\lambda'$  decreasing for the accretion dominated halos in agreement 
with VK01. \label{sparevolall} } 
\end{figure}

We find this to be valid for both spin parameters with a nevertheless 
important and surprising difference. In Fig.\ref{sparevolall} we see that 
$\ovl'$, averaged over {\it all} halos is approximately constant between 
$z=0$ and $z\approx 2$. The mean value of the undashed classical $\lambda$ 
in contrast is significantly increasing with cosmological time, following 
approximately the linear relation
\beq
\lzm=0.039-3.5\times 10^{-3}z.
\eeq

It is instructive to consider just the mM/accretion dominated subsample of halos. 
First we note that the values of both $\lambda$ and $\lambda'$ lie well below 
the full sample of (MM+mM) mean values as a result of the missing angular 
momentum ``kicks''
by major mergers. But while $\oli\lambda'$ due to mMs decreases significantly
with decreasing redshift, we find that the opposite is true for the undashed
classical $\oli\lambda$. This unveils the crucial difference between both
spin parameters. 

To understand this feature we consider again 
Fig.\ref{sratiodist} which shows the spin transfer distributions.
As already mentioned, one would hardly speak of distinct behaviours 
between the transfer rates at first glance: We noted values of $\oli\tl=1.007$ 
and $\oli\tl'=0.997$ for the mM halos. In our simulation every halo suffers 
this rise (or reduction) of its
spin parameter around 18-19 times\footnote{Remember that we defined minor
mergers/accretion events as any mass acquisition between two data outputs that
can not be classified as major merger. The latter happens about once during a 
typical halo life. Since we have 20 output dumps since $z=2$, in 19 of them we 
have to deal with mMs by definition} on average between $z=2$ and $z=0$. 
The net effect of all individual mM events is thus
$\Delta\lp:=\lp_{z=0}/\lp_{z=2}=[\tl^{(\prime)}]^{19}$. For the dashed  
version this gives $\Delta\lambda'=0.997^{19}\approx0.94$, while for the
classic version we get $\Delta\lambda=1.007^{19}\approx1.14$. 
Fig.\ref{sparevolall} shows that these numbers correspond roughly to the
redshift evolution of either minor merger sample! Thus the tiny differences
between the $\tl$ and $\tl'$ distributions lead to a distinct 
time evolution of the spin parameters. 
It is beyond the scope of this paper to analyse the detailed physical 
mechanisms for this difference.

According to the  Fig.\ref{sparevolall}, major mergers in either case 
have only corrective character: Though individual MMs rise the spin parameter 
by 130\% on average, they occur too rarely to overtrump the net effect of
accretion and minor mergers which act quasi permanently.  We conclude 
that minor merger events and accretion rather than major mergers provide the 
driving force guiding the behaviour of the spin parameters. Their very 
distinct time evolution is entirely caused by tiny differences of the spin 
parameters' response to accretion, while major mergers yield a minor additional 
effect, acting similarly on either $\lambda$ and $\lambda'$. 
Minor mergers continuously bring in a net amount of angular momentum.
Large scale gravitational torques therefore seem to dominate the origin
of angular momentum in galaxies in conflict with previous ideas that
continuous infall occurs from random directions, leading to 
a zero net effect on $\lp$.  

\section{Summary and Discussion }

We performed simulations of large scale structure formation in a $\Lambda$CDM
universe and extracted all {\it Friends-of-Friends} selected dark matter halos 
at 20 distinct
redshifts between $z=2.8$ and $z=0$. Our principal goal was to investigate 
the differences of the two commonly used versions of the spin parameter, to understand
their behaviour under the influence of halo major mergers and 
accretion and to determine whether the spin parameter evolution is 
driven by minor ao major mergers. 


The spin transfer $\tlp=\lf^{(\prime)}/\li^{(\prime)}$ for 
any minor (mM) and major merger (MM) event throughout the simulations was 
calculated . We find that (1) the probability distribution $\tl'$ agrees well 
with the one found by Vitvitska et al. (2004) and (2) the corresponding ``classical'' 
quantity $\tl$ is distributed almost identically. This holds for either minor
and major mergers regardless of the considerable distinct values of 
$\lambda$ and $\lambda'$ especially during merger events. 

We analysed the spin transfer dependence of the orbital parameters. The 
pericenter distance between a halo and an infalling satellite was 
identified as the free parameter which is mainly responsible for the spin 
transfer $\tlp$. 
This is a result of the high 
values of $L_{\rm orb}/J_{\rm int}>3 (5.7,11)$  in 80\% (50\%, 20\%) of all 
major mergers ($L_{\rm orb}$ and $J_{\rm int}$ being the orbital and internal 
angular momentum, respectively). It is mainly orbital angular momentum that is 
transferred to internal spin during mergers. The relative orientations 
between halo and satellite spin vectors turned out to have a minor (but 
detectable) influence on $\tlp$, resulting from similar order-of-magnitude 
values of internal and orbital angular momentum in a considerable amount of 
merger events. 

In order to further analyse the effect of merger events on the evolution
of the spin parameters, we plotted the evolution of $\lp$ along a fictitious 
time axis which describes the time relative to the final MM event of each
individual halo. It turned out that after the merger onset, $\lp$ 
experiences a peaked rise of $\approx 30\%$, and a considerable decline
later on. The first is due to relaxation effects, combined with the transformation 
of orbital to internal angular momentum, while the latter seems to be  the result of 
subsequent mass accretion in combination with a constant mean value of the 
added specific angular momentum. 

The time evolution of the virial coefficient $\eta=2\ek/|\ep|$ and its 
correlation with the spin parameters was investigated.  
$\lp$ is very sensitive to $\eta$ following $\lp\approx\alpha+\beta\eta^4$. 
The values of $\alpha$ and $\beta$ depend on redshift and the type
of the spin parameter. The correlation is stronger for lower redshifts and
for the dashed spin parameter $\lambda'$. On the fictitious time axis $t^*$, 
$\eta$ shows the same peak-like feature around $t^*=0$ as the spin parameters, 
though less pronounced. A strong correlation was found between the ratio
$\lambda'/\lambda$ and $\eta$. While one might not be surprised about a 
$\lp-\eta$
correlation in principal (due to the direct or indirect use of the mechanical
energies) the detailed origin of this particular behaviour is still an open 
issue. 

We finally examined the redshift evolution of the spin parameters. 
While the global mean value $\oli\lambda'$ turnes out to be roughly redshift 
independent, we find that $\oli\lambda$ is significantly increasing with time.
If we restrict the halo sample to those which never had major mergers, the
picture changes slightly: While $\lambda'$ is still significantly growing
with time, we find $\lambda$ to decrease moderately. The earlier result of 
similar $\tlp$ distributions is not contradicting this result: In fact 
$\lambda$ and $\lambda'$ have mean spin transfers which are different by
a small but important amount
for the mM sample of halos ($\oli\tlp=1.007 (0.997)$). This causes a
significant difference in the global evolution since the ``operation''
$\lp\longrightarrow \tlp\times\lp$ is acting about 18 times on each halo 
between $z=2.8$ and $z=0$. The distinct behaviour of both spin parameters is
thus induced by their distinct reactions on minor merger and accretion events.
Due to their comparable rare appearance major mergers must be considered as 
only minor additional corrections to the $\lambda$ evolution which is
dominated by a coordinated action of minor mergers caused probably by large
scale stream flows that could be the result of tidal torques. 
\newline

We like to thank Hans-Walter Rix for enriching comments and general support. 
This work was supported by the {\it Deutsche Forschungsgemeinschaft} via
{\it SFB 439, Galaxien im jungen Universum}.

\label{lastpage}

\end{document}